\documentclass{iau}

\usepackage{amsmath}
\usepackage{graphicx}
\usepackage{multirow}

\newcommand{\iram}{IRAM-30\,m}

\newcommand{\farc}{\mbox{$.\!\!^{\prime\prime}$}}
\newcommand{\arcmin}{\mbox{$^{\prime}$}}
\newcommand{\fs}{\mbox{$.\!\!^{\rm s}$}}
\newcommand{\mydeg}{$^{\circ }$}

\newcommand{\mloss}{\mbox{$\dot{M}$}}

\newcommand{\my}{\mbox{$M_{\odot}$~yr$^{-1}$}}

\newcommand{\kms}{\mbox{km\,s$^{-1}$}}

\newcommand{\vlsr}{\mbox{$V_{\rm LSR}$}}

\newcommand{\h}{$^{\rm h}$}
\newcommand{\m}{$^{\rm m}$}

\newcommand{\dense}{\mbox{$n_{\rm e}$}}

\newcommand{\htal}{\mbox{H30$\alpha$}}
\newcommand{\htnal}{\mbox{H39$\alpha$}}

\def\snu#1{\ifmmode {S_\nu\,\propto\,\nu^{#1}}
          \else \hbox{$S_\nu$\,$\propto$\,$\nu^{#1}$}\fi}
\def\cm#1{\ifmmode {\,{\rm cm^{-#1}}}                  
        \else \hbox{$\,${\rm cm$^{\rm -#1}$}}\fi}
\def\raw {\ifmmode\rightarrow\else$\rightarrow$\fi}
\def\ex#1{\ifmmode {\times 10^{#1}}         
        \else \hbox{{$\times 10^{\rm #1}$}}\fi}

\begin{document}

\lefttitle{Sánchez Contreras et al.}
\righttitle{Zooming on the emerging ionized regions of pPNe with ALMA}

\journaltitle{Planetary Nebulae: a Universal Toolbox in the Era of Precision Astrophysics}
\jnlDoiYr{2023}
\doival{10.1017/xxxxx}
\volno{384}

\aopheadtitle{Proceedings IAU Symposium}
\editors{O. De Marco, A. Zijlstra, R. Szczerba, eds.}
 
\title{Zooming on the emerging ionized regions of pPNe with ALMA}

\author{C. Sánchez Contreras$^1$, D. Tafoya$^2$, J. P. Fonfr{\'i}a$^1$, J. Alcolea$^3$, A. Castro-Carrizo$^4$, and V. Bujarrabal$^3$}
\affiliation{$^1$ Centro de Astrobiolog{\'i}a (CAB), CSIC-INTA. 
Postal address: ESAC, Camino Bajo del Castillo s/n, E-28692, 
Villanueva de la Ca\~nada, Madrid, Spain.}
\affiliation{$^2$ Department of Space, Earth and Environment, Chalmers University of Technology, Onsala Space Observatory, 439 92 Onsala, Sweden.}
\affiliation{$^3$Observatorio Astron\'omico Nacional (IGN), Alfonso XII No 3, 28014 Madrid, Spain.}
\affiliation{$^4$Institut de Radioastronomie Millimetrique, 300 rue de la Piscine, 38406 Saint Martin d’Heres, France.}

\begin{abstract}
We report on recent results from our successful and pioneering
observational program with ALMA to study emerging ultracompact HII
regions of pre-planetary nebulae (pPNe) using mm-wavelength
recombination lines (mRRLs) as new optimal tracers. We focus on our
study of two poster-child pPNe, namely, M\,2-9 and CRL\,618. We reveal
the structure and kinematics of the enigmatic inner nebular regions of
these objects with an unprecedented angular resolution down to 20-30\,mas
($\sim$15-30\,AU linear scales).  For both targets, the
ionized central regions are elongated along the main symmetry axis of
the large-scale nebulae, consistent with bipolar winds, and show
notable axial velocity gradients with expansion velocities of up to
$\sim$100\,\kms. The intensity and width of the H30$\alpha$
profiles are found to be time variable, denoting changes on scales of
a few years of the physical properties and kinematics of the present-day 
post-AGB ejections. Our ongoing analysis involves 3D, non-LTE radiative
transfer modeling of the mRRLs and free-free continuum
emission. This approach allows us to provide an exceptionally detailed
description of the physical conditions in the innermost layers of
these well known pPNe.

\end{abstract}

\begin{keywords}
stars: AGB and post-AGB, circumstellar matter, stars: mass-loss,
stars: winds, outflows, planetary nebulae: individual (M\,2-9, CRL\,618).
\end{keywords}

\maketitle

\section{Introduction}
Previous studies of pPNe and young PNe (yPNe) suggest that multiple lobes and high
velocities may result from collimated fast winds (CFWs or ``jets")
interacting with slowly expanding circumstellar envelopes formed
during the previous AGB phase \citep[see e.g.\,][for a comprehensive
  review]{bal02}. However, directly characterizing post-AGB jets and
their launch regions within a few hundred astronomical units is
challenging due to their small angular sizes and significant
obscuration caused by optically thick circumstellar dust shells or
disks.
A recent pilot study using millimeter radio recombination lines
(mRRLs) in a sample of pPNe/yPNe observed with the
\iram\ radiotelescope shows that mRRLs are optimal tracers to probe
very deep ionized nebular regions ($\lsim$150 au), taking us closer to the
sites where CFWs are launched and facilitating the measurement of the
mass-loss rate of ongoing post-AGB ejections \citep[][hereafter
  CSC+17]{san17}. CSC+17 reported the detection of mRRLs in three
(out of eight) objects within their sample: MWC\,922, M\,2-9, and
CRL\,618, resulting in the first characterization of the
spatio-kinematic structure and physical conditions of their central
ionized regions.

For M2-9 and CRL\,618, the data revealed the existence of young
($\lsim$15-20\,yr) bipolar outflows characterized by moderate
average velocities (of a few tens of \kms) and relatively high mass-loss rates
($\approx$$10^{-7}$-$10^{-6}$\,\my).  In the case of MWC\,922,
an IR excess B[e]-type star surrounded by a large-scale
reflection nebulosity displaying a remarkable X-shape morphology, 
the line profiles suggested the presence of both a rotating disk and a
wind. Through a subsequent ALMA study, \cite{san19} provided
evidence for a spatially resolved, nearly edge-on rotating disk and
also uncovered a fast ($\sim$100\,\kms) bipolar wind perpendicular to
the disk, which was found to rotate in tandem with the disk itself.

This contribution outlines our ongoing ALMA-based study of the
emerging compact ionized region at the cores of M\,2-9 and
CRL\,618.

\section{Observations}
\label{obs}
The observations of M\,2-9 and CRL\,618 were conducted using the ALMA
12-m array as part of projects 2016.1.00161.S and 2017.1.00376.S. The
observations were carried out in Band 3 (3\,mm) and Band 6 (1\,mm),
with a total of twelve different spectral windows (SPWs) dedicated to
mapping the emission of various mRRLs (and CO lines) as well as the
continuum.  Observations were conducted in October and November 2017
using 45-50 antennas, with baselines ranging from 41.4\,m to 16.2\,km
for Band 3, and from 41.4\,m to 14.9\,km for Band 6.  The maximum
recoverable scale (MRS) of the observations is $\sim$0\farc8 and
$\sim$0\farc7 at 3 and 1\,mm, respectively.  Full details on the
observations and imaging techniques for M\,2-9 and CRL\,618
can be found in S\'anchez Contreras et al. and Fonfr{\'i}a et al., in
preparation, respectively.  The final mRRLs cubes and continuum maps
presented here have angular resolution of $\sim$40-60\,mas at
1\,mm and $\sim$70-90\,mas at 3\,mm.

\begin{figure}
\begin{center}
\includegraphics[width=0.45\hsize]{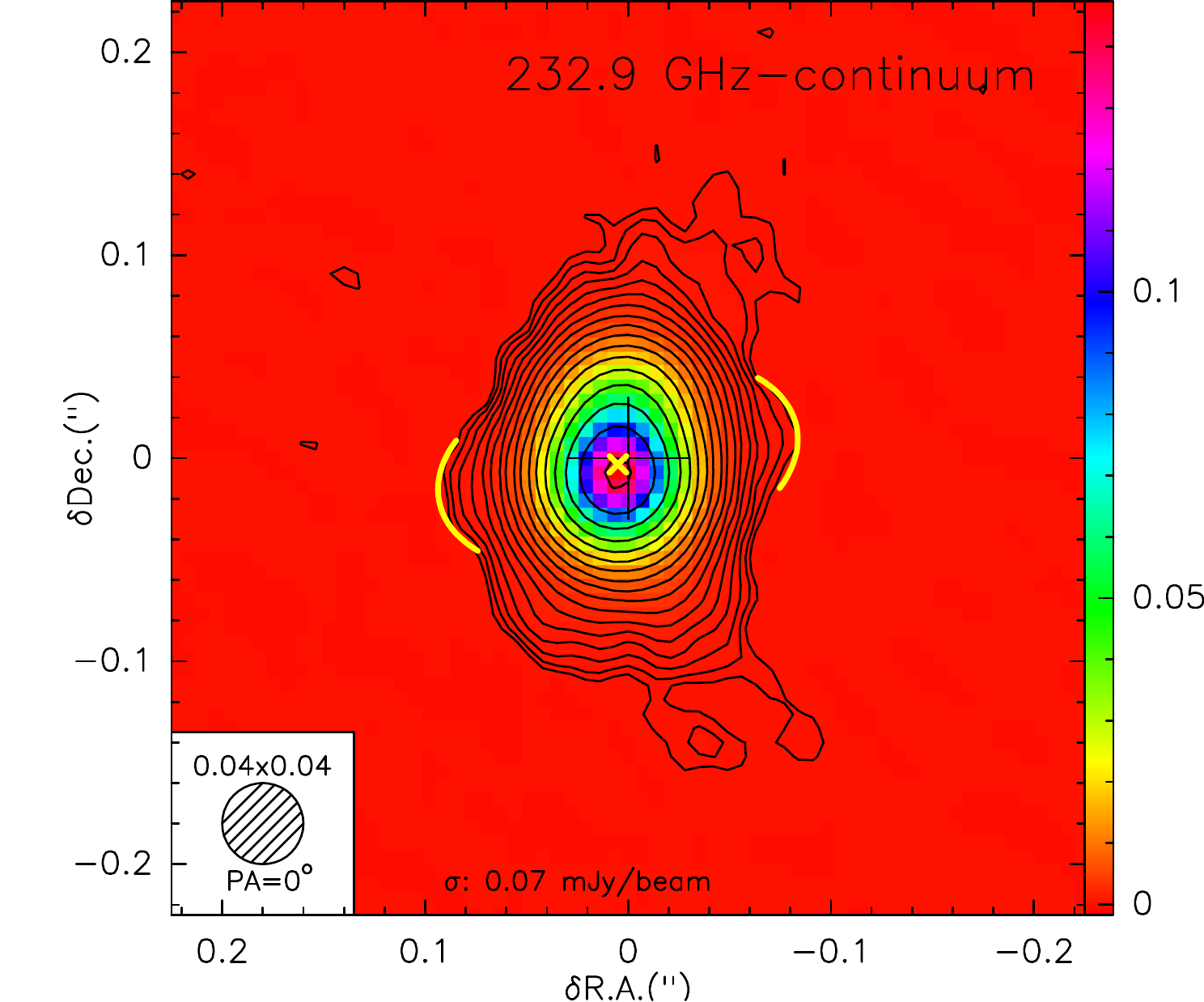}
\includegraphics[width=0.45\hsize]{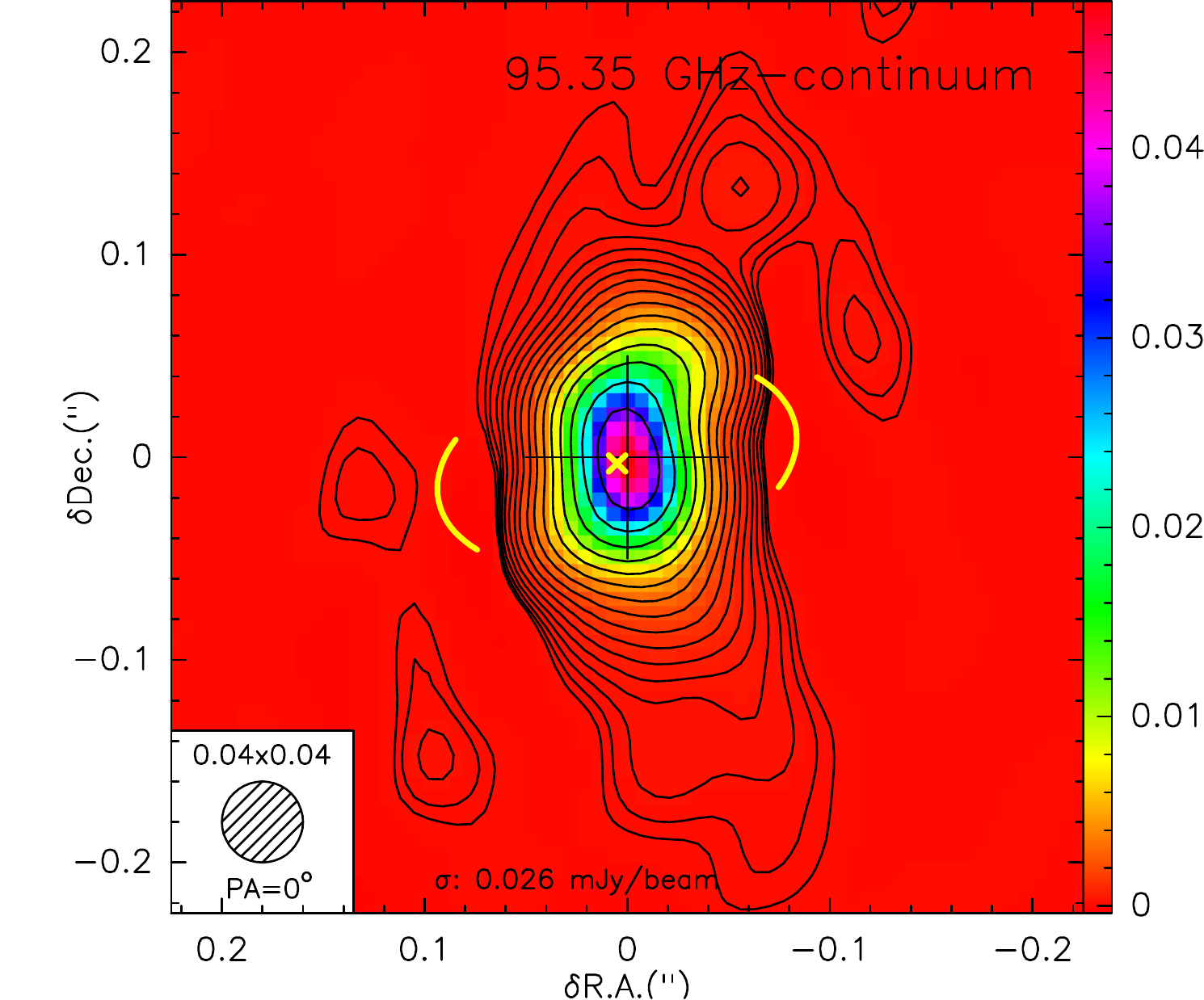}   
\end{center}
\caption{ALMA continuum emission maps of M\,2-9 at 232.9\,GHz
     (left) and 95.4\,GHz (right) using a circular restoring beam with half power beam width of HPBW=0\farc04. The level
     contours are (3$\sigma$)$\times$1.5$^{(i-1)}$\,Jy\,beam$^{-1}$,
     $i$=1,2,3...  The central cross marks the 3\,mm continuum surface
     brightness peak at coordinates J2000 R.A.=17\h05\m37\fs96679 and
     Dec.=$-$10\mydeg08\arcmin32\farc65 (J2000). The yellow arcs, centered at the small yellow cross, 
     represent the broad-waist structure, plausibly a dust disc.}
\label{cont-m29} 
\end{figure}

\section{M\,2-9 or ``The Butterfly Nebula''}

M\,2-9 is a very well-documented yPNe candidate that has gathered
significant attention in the extisting literature \citep[e.g.][and
  references therein]{kwo85,cor11,cc12,cc17,san17,bal18}. It has
prominent large-scale lobes (oriented along the NS direction) with a
dominant expansive kinematics.  Indirect, yet compelling evidence
suggests that M\,2-9 hosts a binary system with an orbital period of
$\sim$90\,yr, although the nature of this binary remains unknown. The
extended lobes display observable radio-continuum free-free emission,
contributing to a nearly flat spectral energy distribution at
centimeter wavelengths. Additionally, a compact ($\lsim$0\farc2)
region at the core emits radio-continuum, inferred to be an ionized
wind based on the spectral index (\snu{(0.6-0.8)}), which has been mapped
using ALMA as part of this project.

The distance to M\,2-9 is highly uncertain, with values spanning from
50\,pc to 3\,kpc in the literature.  The two most accepted values are
$d$=1.3\,kpc \citep{cor11} and $d$=650\,pc \citep{cc12} from
the analysis of proper motions and other nebular properties 
at optical and mm wavelengths, respectively.  We adopt a value of
$d$=650\,pc, which allows direct comparison with the study by CSC+17.

\subsection{An ionized bent jet surrounded by a circumbinary disk}

The ALMA continuum maps at 1 and 3\,mm (Fig.\,\ref{cont-m29}), show an elongated structure that resembles what is
seen at longer wavelengths. At 3\,mm, we discern a C-shaped curvature similar
to that observed in 2015 in the VLA 7\,mm maps \citep{dlF22} consistent
with a bent collimated wind or jet emerging from the core. This wind has dimensions
of 0\farc4$\times$0\farc13 (260$\times$85\,au at $d$=650\,pc) at 3\,mm.
We measure a spectral index of the continuum of \snu{0.92\pm0.10}, which 
  suggests predominantly free-free emission from the ionized jet, with
  a minor contribution from dust. 
  
There are some differences between the 1\,mm and 3\,mm maps that partly arise
from the frequency-dependence of the free-free continuum optical depth and from
the increased contribution of thermal emission from dust to the
continuum at shorter wavelengths. In particular, the 1\,mm maps show a
broad-waist of emission, absent at 3\,mm and in the mRRLs maps, which
very likely represents a dusty equatorial disk (with a radius of
$\sim$50\,au) surrounding the ionized jet. This mm-continum disk is
probably the counterpart of the compact, dust disk known to exist at
the core of M\,2-9 based on mid-infrarred observations \citep{lyk11}.

Further support for the existence of an equatorial disk around the ionized jet 
comes from our observation of CO compact absorption at the
center (not shown). Notably, this absorption is slightly offset to the north (by
$\sim$0\farc012). Considering the system's inclination, with the
north lobe oriented at an angle of $i$$\sim$17\mydeg\ away from the
plane of the sky, this offset is in line with the absorption of the
background free-free continuum (from the ionized core) occurring by
the front part of an equatorial disk. The CO absorption feature is
redshifted by $\sim$6\,\kms\ from the centroid of the mRRLs, which is
most likely due to infall motions from the inner regions of the circumbinary 
disk.

\subsection{Kinematics of the ionized jet}

\begin{figure}
\begin{center}
     \includegraphics*[width=0.28\hsize]{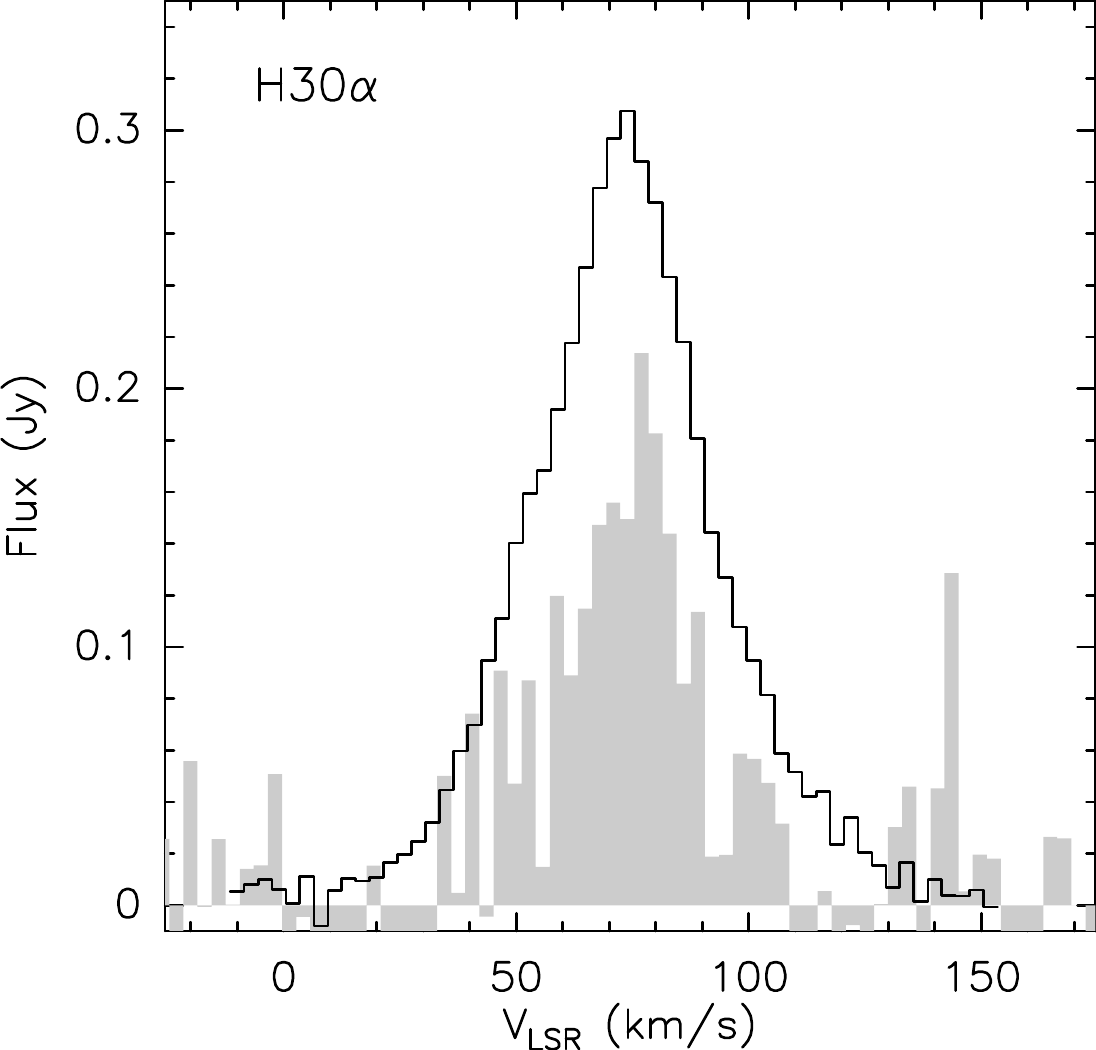}                
     \includegraphics*[width=0.32\hsize]{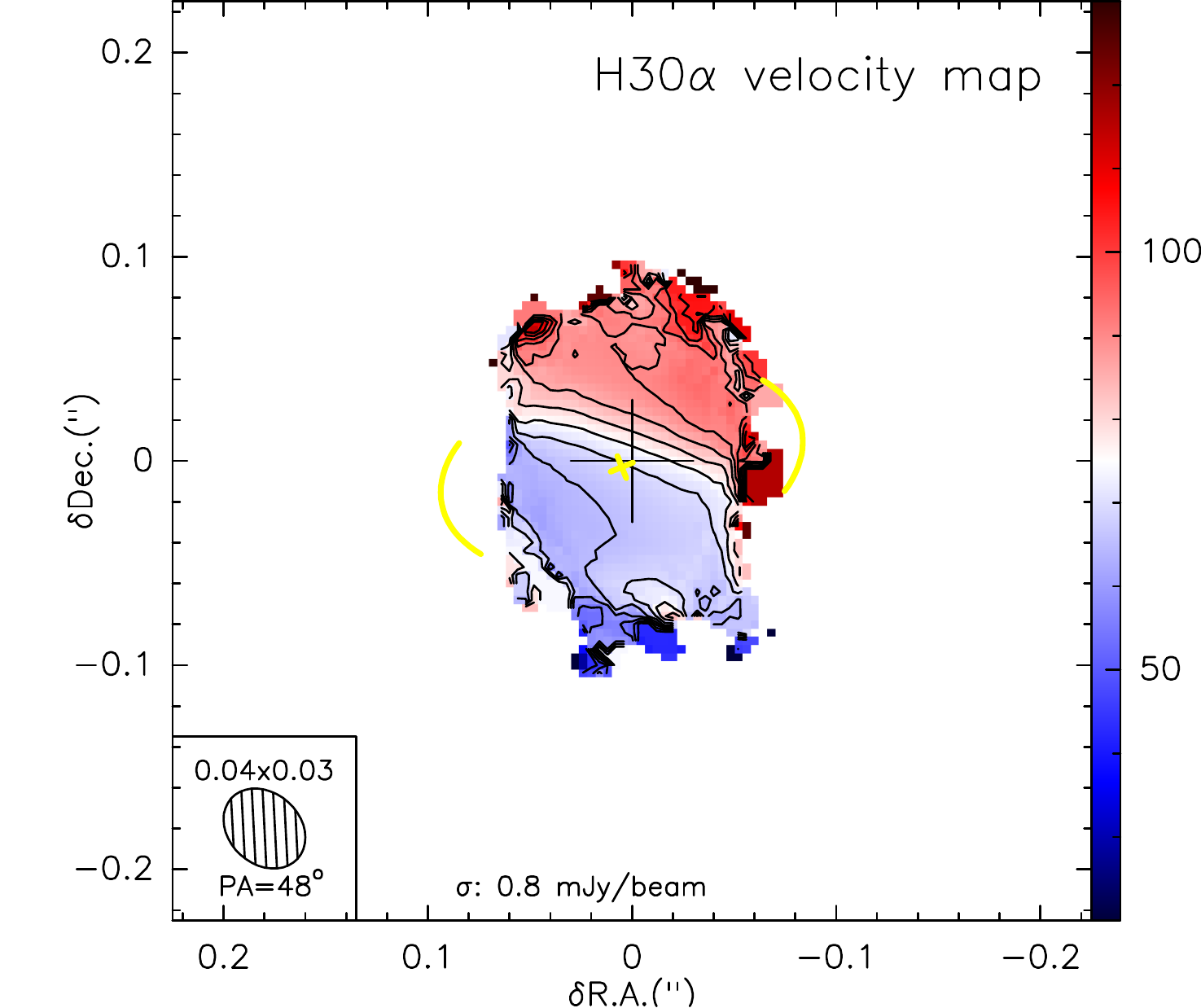}               
     \includegraphics*[width=0.38\hsize]{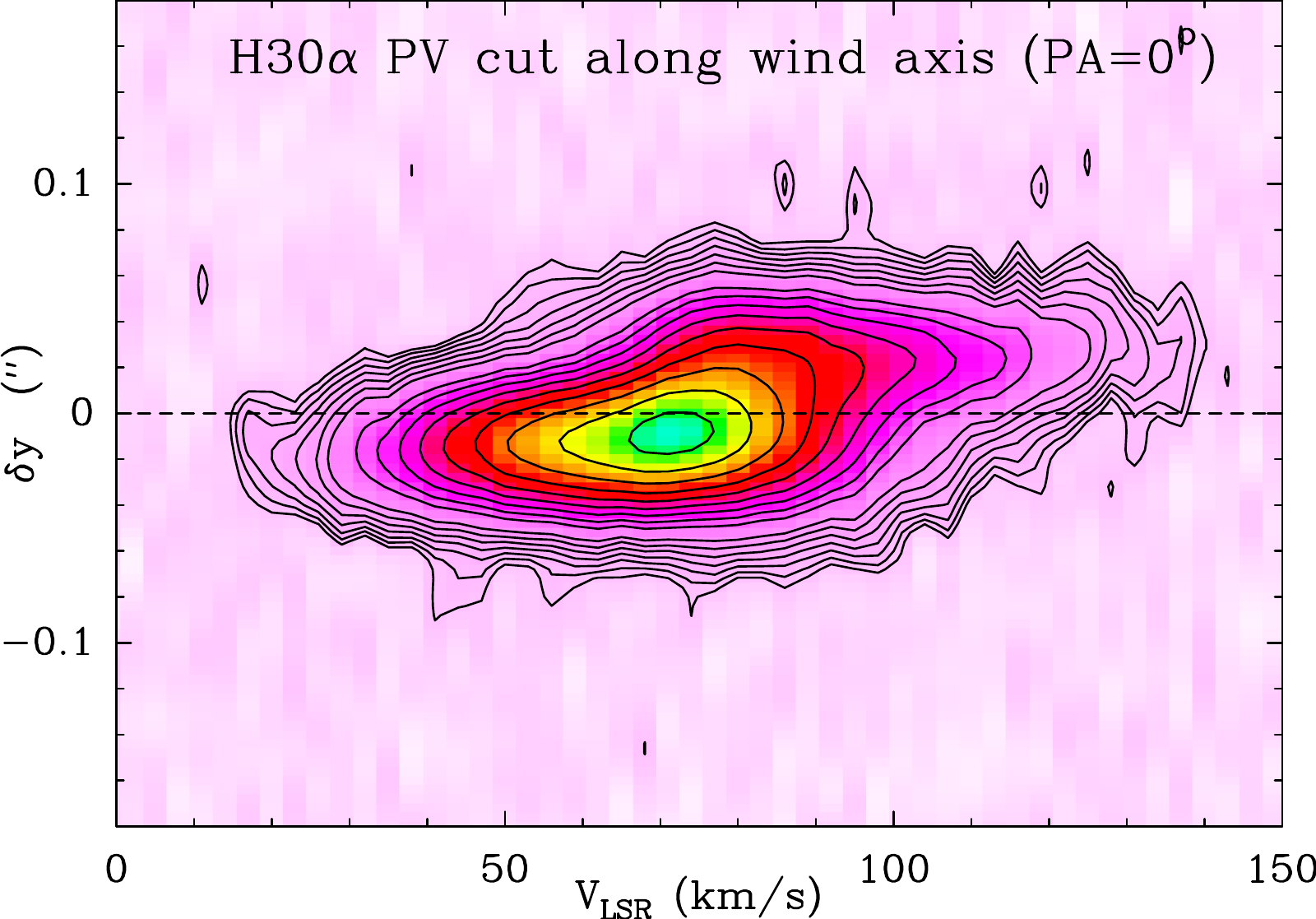} 
     \\
     \includegraphics*[width=0.28\hsize]{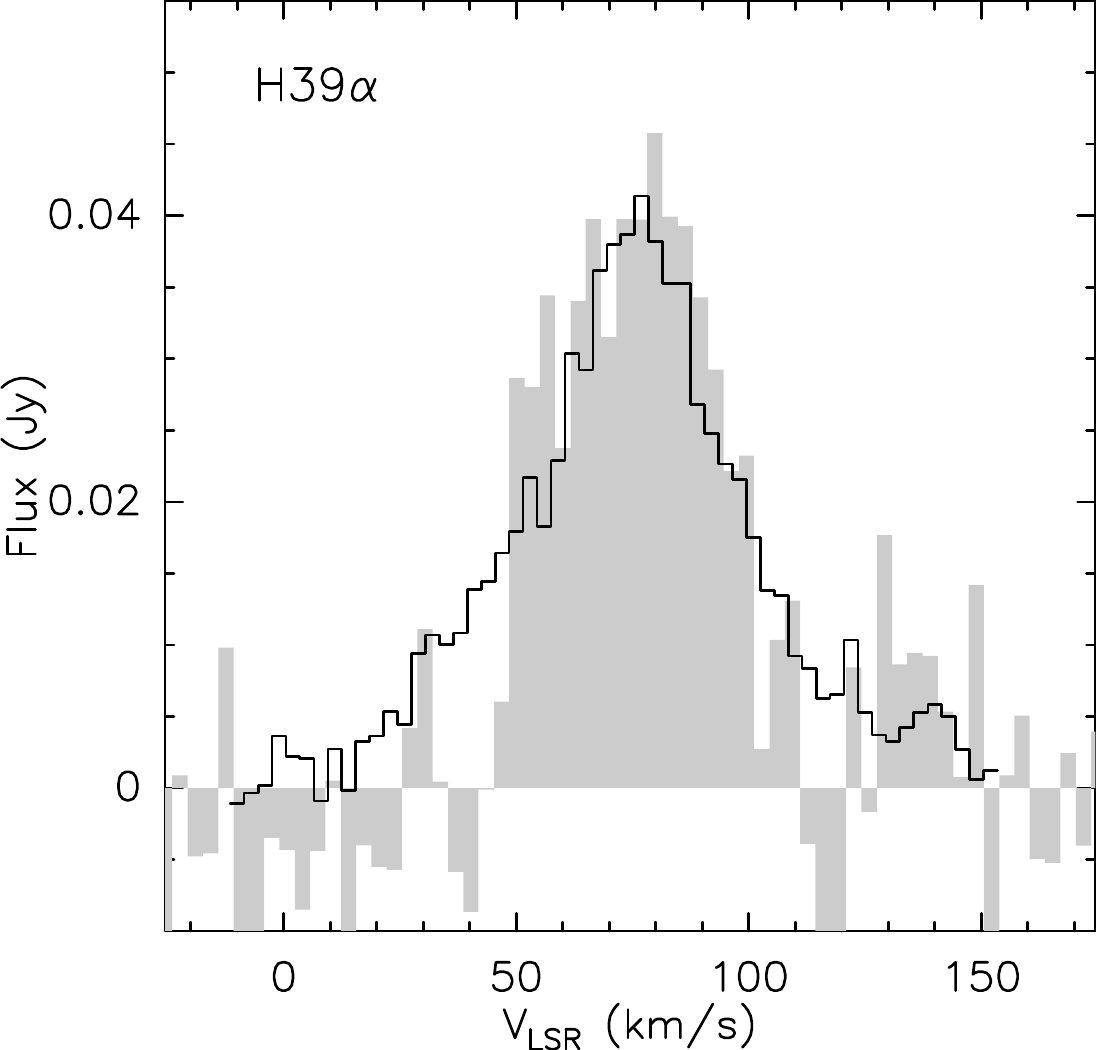}                 
     \includegraphics*[width=0.32\hsize]{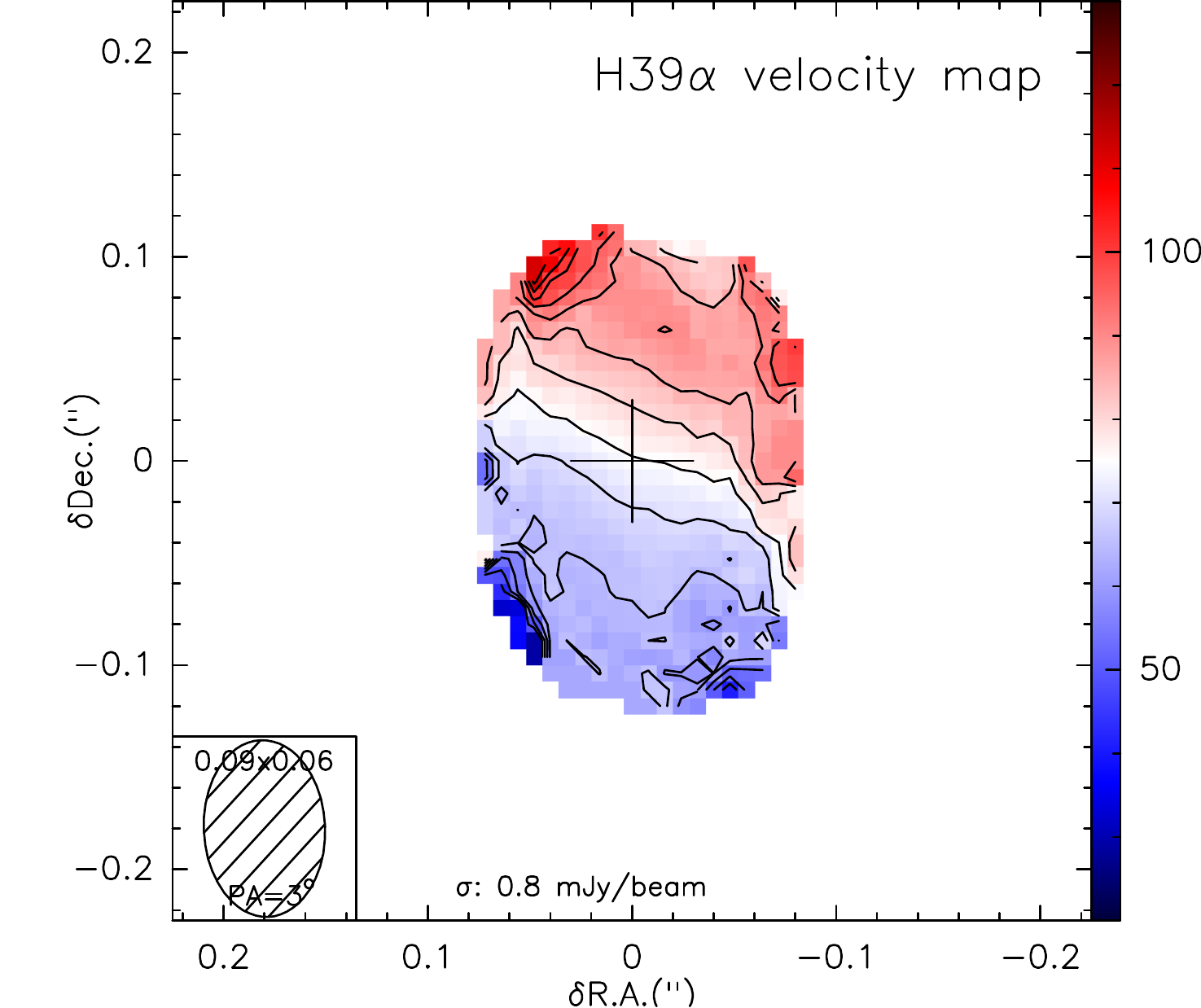}               
     \includegraphics*[width=0.38\hsize]{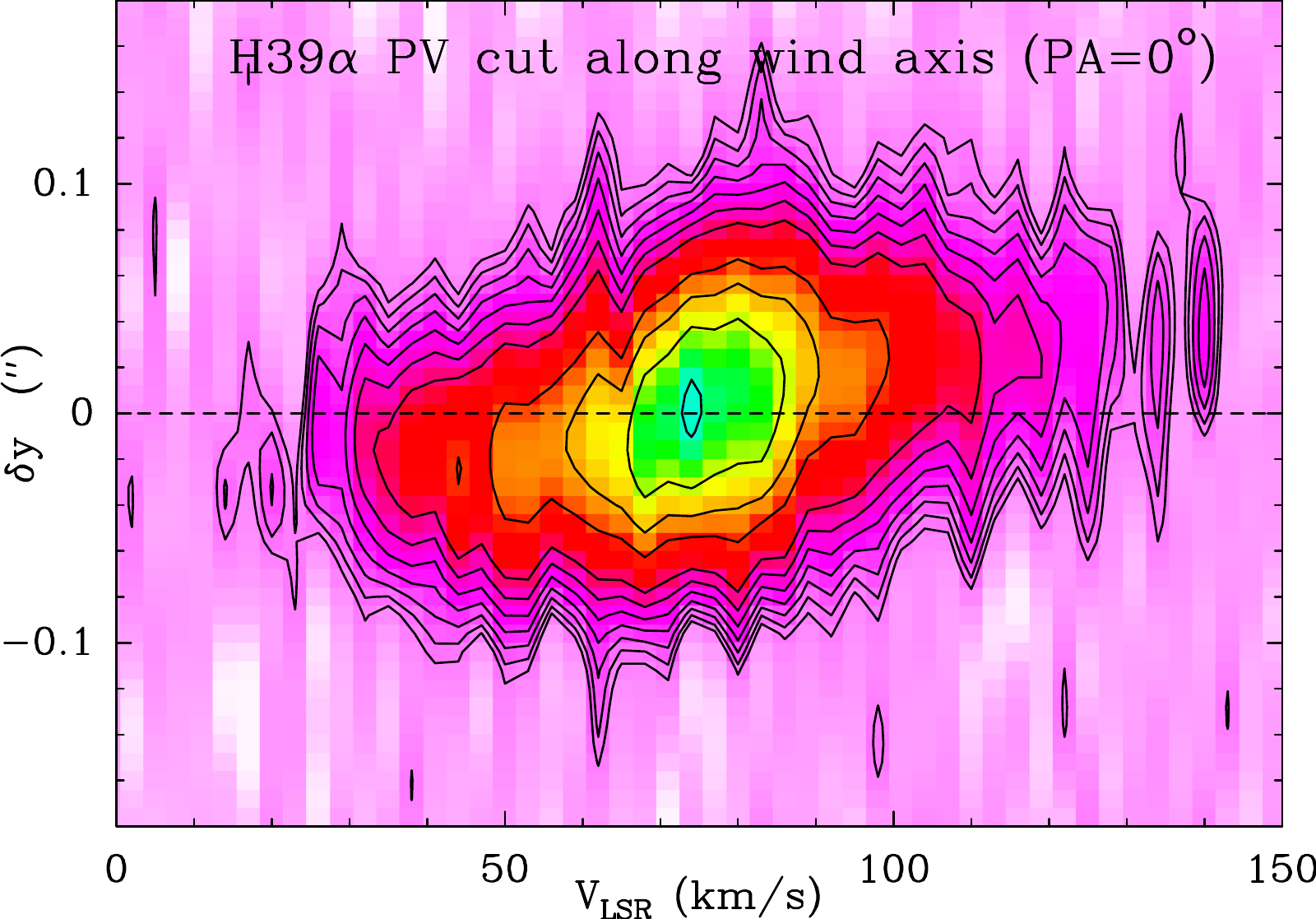} 
\end{center}
\caption{Summary of ALMA data of the \htal\ (top) and \htnal\ (bottom) recombination lines. Left: Integrated line spectrum
           obtained with ALMA (black lines) and
           with the \iram\ antenna (grey histogram, CSC+17). 
             Middle: First moment map. Contours going from \vlsr=45 to 115\,\kms\ by 5\,\kms. The wedge indicates the \vlsr-color relationship.
             Right: Position velocity cuts through the center along the wind axis (PA=0\mydeg). Levels are 2.5$\times$(1.3)$^{(i-1)}$ for \htal\ and 1.5$\times$(1.3)$^{(i-1)}$ for \htnal\ with $i$=1,2,3... 
  }
\label{mRRLs-m29} 
\end{figure}

As shown in Fig.\,\ref{mRRLs-m29}, the \htal\ and \htnal\ line
emission span a wide velocity range of $\sim$140\,\kms, sharing a
similar extent and morphology with the mm continuum emission (with the
exception of the broad waist at 1\,mm). The comparison between the
ALMA line profiles (integrated over the emitting area) with those
obtained 2 years earlier with the \iram\ antenna (CSC+17) shows 
that the lines are now more intense and broader, implying
changes of in the wind properties over yearly time scales.

The ionized jet's kinematics can be examined using the moment-1 maps
and position-velocity diagrams. We observe a
global expansion indicated by an overall velocity gradient along the
nebula's symmetry axis (PA=0\mydeg). The velocity gradient suggests
very short kinematical ages of less than approximately one year,
consistent with the observed changes in the line profiles over the
course of a year. Additionally, we note a slight velocity gradient
perpendicular to the lobes, which is evident in the sloping (non-horizontal)
isovelocity contours in the momentum-1 map, tentatively suggesting
jet rotation. The axial position-velocity diagrams display a
distinctive S-shape, showing maximum line widths at two compact
regions located diametrically opposed along the axis (at offsets of
$\sim$20-30\,mas), indicating either rapid/abrupt wind acceleration or
shocks at these compact regions (referred to as $HVspots$).

\subsection{Non-LTE radiative transfer modeling}

For a comprehensive analysis of the spatio-kinematics and physical
conditions of the ionized jet, we are currently conducting radiative
transfer modeling using the co3Ral code developed by co-I D.\,Tafoya
(S\'anchez Contreras et al.\, in prep.). Although the modeling is
ongoing, it appears that a wind with a non-uniform density structure
is necessary to account for the observations. Notably, the HVspots are
well depicted by two regions of high-density and high-velocity,
resembling the shock-compressed bipolar structure proposed by \cite{liv01} in
their model for M\,2-9, which is based on the interaction of
a tenuous companion-launched jet and the dense
primary star's wind.
We find an average electron temperature of
$\sim$15000\,K and electron densities ranging from
\dense$\approx$10$^6$ to 10$^9$\,\cm3. The average mass-loss rate of
the ionized wind is deduced to be of \mloss$\approx$10$^{-7}$\,\my.

\section{CRL\,618 or ``The Westbrook Nebula''}

\begin{figure}[htp!]
  \begin{center}
 \includegraphics[width=0.99\hsize]{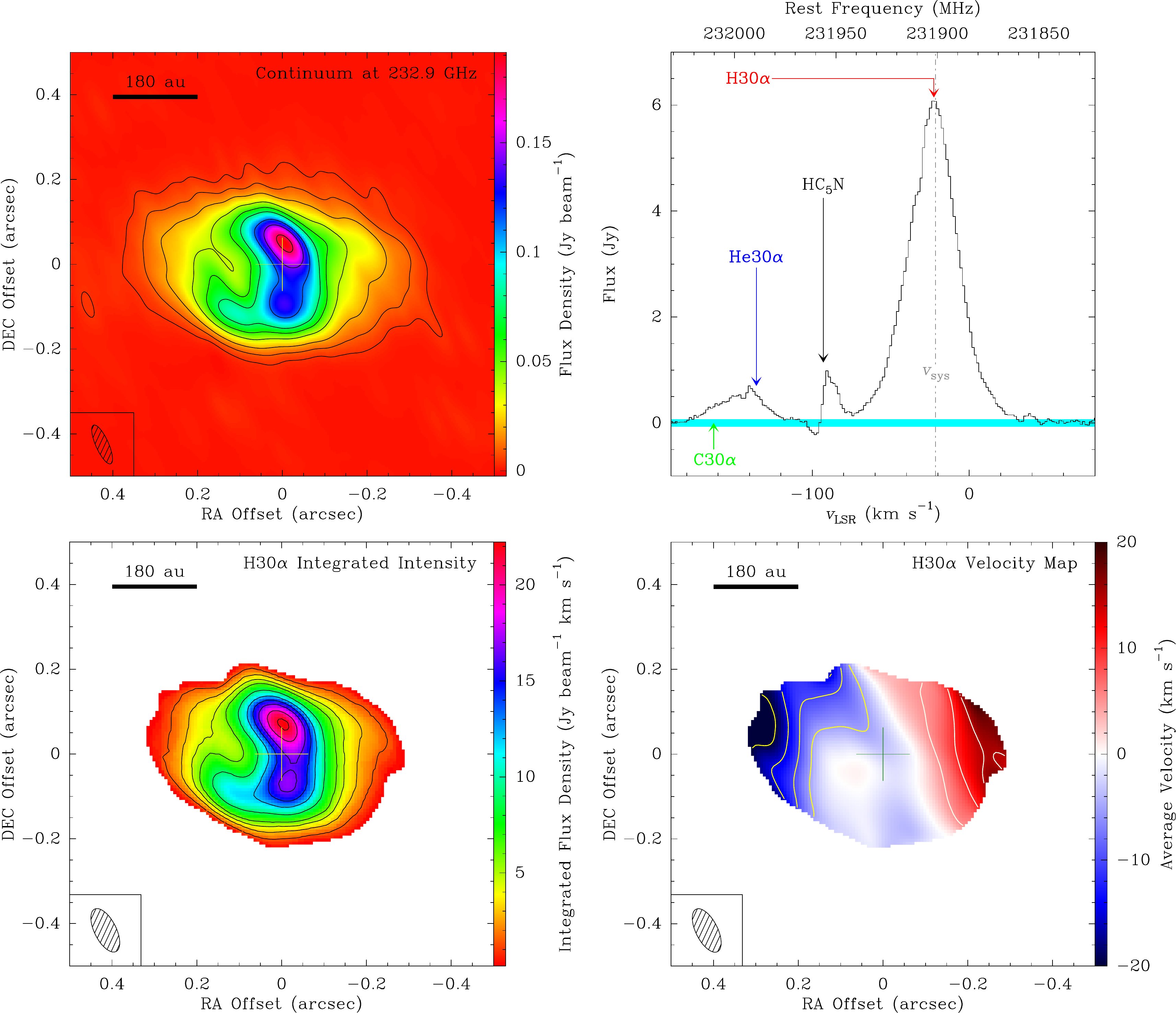}
  \end{center}
  \caption{Summary of ALMA data for CRL\,618. Top-left: Continuum
    emission map at 232.9\,GHz.  Top-right: Spectrum of
    \htal\ integrated within the emitting area (other lines are also
    labeled).  Bottom: Zeroth and first moment maps (left and right,
    respectively) of \htal.  The ellipse at the bottom-left corner of
    the maps represents the HPBW. The central cross marks the
    coordinates of the tracking center at R.A.= 04\h42\m53\fs583 and
    Dec.= 36\mydeg06\arcmin53\farc34 (J2000).  }
\label{f-c618} 
\end{figure}

CRL\,618 is a well-studied pPN exhibiting a complex and intricate
multipolar structure, including fast outflows reaching expansion
velocities of up to $\sim$200\,\kms\ and predominantely oriented in the east-west direction
\citep[e.g.][]{cer89,san02,san04a,rie14}.  Compared with
M2-9, CRL\,618 has a younger and denser envelope, with the majority of
its mass existing in the form of molecular gas \citep{san04b,par07}. A small fraction of
material at its core gets ionized by a central B-type star, resulting
in the emission of free-free continuum radiation and mRRLs \citep{kwo81,mar88,taf13}. These
mRRLs were initially observed by \cite{mar88} and revisited by us in 2015 
using the \iram\ antenna (CSC+17), uncovering significant
changes in their profiles and confirming ongoing fluctuations in
free-free continuum emission over the past few decades.

We observed with ALMA the central ionized region of CRL\,618 in 2017
within projects 2016.1.00161.S and 2017.1.00376.S. Regrettably, in
this case, the 3\,mm data failed the quality assessment, leaving us
only with the band 6 data (Fig.\,\ref{f-c618}).  Similarly to M\,2-9, the 1\,mm continuum
emission exhibits an elongation aligned with the main nebula axis,  
mirroring patterns observed at cm wavelengths, albeit on a smaller scale due to the lower opacity of the
free-free continuum at mm-wavelengths. The brightness distribution is
non-uniform and suggests a hollow cylindrical geometry, with inner and outer radius of
$\sim$70 and 110\,au, respectively, with a dense
equatorial zone. 

The \htal\ line profile exhibits broad wings, implying gas movements
at speeds of $\sim$70-100 km/s within the cylindrical shell emitting
the free-free continuum. The velocity map reveals a gradient along the
east-west direction, i.e., the nebula main symmetry axis, indicating
expansion and resulting in kinematical ages of just a few
years. Despite the seemingly smooth appearance of the velocity map, a
closer examination of the position-velocity diagrams indicates
underlying complexity that we are currently investigating.  For
instance, the velocity field might encompass radial and axial
expansions in varying proportions for the equator and the
lobes. Additionally, we have noticed potential asymmetries in the
geometry, such as the west lobe appearing shorter, and the cylindrical
shell possibly being incomplete. We are currently exploring these and
other intricacies using a 3D non-LTE radiative transfer model.
This effort aims to provide the most comprehensive characterization to date of the central regions, within $\sim$300\,au,
of this iconic object.

\section{Concluding remarks}

In this presentation, we underscore the significance of mRRLs in
probing the enigmatic central ionized regions of pPNe/yPNe. These lines offer a unique
opportunity to closely investigate the jet engine and estimate the
mass-loss rate of the ongoing ejections.

In conjunction with the previously published results from our ALMA
project on MWC\,922 \citep{san19}, the current analysis of our ALMA
mRRL emission maps of M\,2-9 and CRL\,618 is delivering a
comprehensive and detailed account of the structure, kinematics, and
physical conditions in the central regions of pPNe. We are observing
diverse properties of the ionized central regions in the three targets observed within our project, emphasizing the
need for additional similar observations to gain meaningful
statistical insights and advance our understanding of the developement of nebular
asymmetries and rapid outflows in these late
evolutionary stages of low-to-intermediate mass stars.

We acknowledge funding from the Spanish MCIN/AEI/10.13039/501100011033
(projects PID2019-105203GB-C22 and PID2019-105203GB-C21). This paper
makes use of the following ALMA data: ADS/JAO\-.ALMA\#2016.1.00161.S and
ADS/JAO.ALMA\#2017.1.00376.S.  ALMA is a partnership of ESO
(representing its member states), NSF (USA) and NINS (Japan), together
with NRC (Canada), MOST and ASIAA (Taiwan), and KASI (Republic of
Korea), in cooperation with the Republic of Chile. The Joint ALMA
Observatory is operated by ESO, AUI/NRAO and NAOJ.

\end{document}